\documentclass[preprint,review,12pt]{elsarticle}

\usepackage{amssymb}
\usepackage{amsmath}
\usepackage{layout}
\usepackage{hyperref}
\usepackage{amsmath,amsfonts}
\usepackage{algorithmic}
\usepackage{array}
\usepackage{textcomp}
\usepackage{stfloats}
\usepackage{url}
\usepackage{verbatim}
\usepackage{graphicx}
\usepackage{graphicx}
\usepackage{float}
\usepackage{geometry}
\usepackage{array}
\usepackage{multirow}
\usepackage{tabularx}
\usepackage{pifont}
\usepackage{booktabs}
\usepackage{graphicx} 
\usepackage{siunitx} 
\usepackage[ruled,vlined]{algorithm2e}
\usepackage{setspace}
\usepackage{pifont} 
\usepackage{caption}
\usepackage{subcaption} 
\journal{Nuclear Physics B}

\begin{document}

\begin{frontmatter}



\title{TenonOS: A Self-Generating LibOS-on-LibOS Framework for Time-Critical Embedded Operating Systems}


\author[1]{Yifan~Zhang}
\author[1]{Xinkui~Zhao\corref{cor1}}
\author[2]{Haidan~Zhao}
\author[2]{Hao~Zhang}
\author[1]{Qingyu~Ma}
\author[4]{Lufei~Zhang}
\author[3]{Guanjie~Cheng}
\author[3]{Shuiguang~Deng}
\author[3]{Jianwei~Yin}
\author[5]{Zuoning~Chen}

\affiliation[1]{organization={College of Software Technology, Zhejiang University},
               city={Ningbo},
               country={China}}

\affiliation[2]{organization={Yingyi Technology Co., Ltd.},
               city={Hangzhou},
               country={China}}

\affiliation[3]{organization={College of Computer Science and Technology, Zhejiang University},
               city={Hangzhou},
               country={China}}

\affiliation[4]{organization={State Key Laboratory of Mathematical Engineering and Advanced Computing},
               city={Wuxi},
               country={China}}

\affiliation[5]{organization={Chinese Academy of Engineering},
               city={Beijing},
               country={China}}

\cortext[cor1]{Corresponding author. Email: \texttt{zhaoxinkui@zju.edu.cn}}
\begin{abstract}
The increasing complexity of modern embedded systems creates a fundamental conflict between the demand for sophisticated functionality and the strict constraints on resources and temporal determinism. Traditional operating system and hypervisor architectures, typically reliant on monolithic designs or rigid abstraction layers, suffer from significant resource bloat and unpredictable scheduling behaviors. These limitations render them ill-suited for time-critical applications where minimizing latency and jitter is paramount.

To address these challenges, we propose TenonOS, a demand-driven, self-generating, and lightweight operating system framework for time-critical embedded systems that fundamentally rethinks and reconstructs both the hypervisor and OS architectures. TenonOS introduces a novel LibOS-on-LibOS architecture, in which both hypervisor and OS functionalities are decomposed into fine-grained, reusable micro-libraries. Unlike static legacy systems, TenonOS employs a generative orchestration engine that dynamically composes these modules to synthesize a customized runtime environment tailored specifically to the application's criticality level, timing constraints, and resource profile.

At the core of this framework are two synergistic components: Mortise, a minimalist micro-hypervisor, and Tenon, a real-time LibOS. Mortise provides lightweight resource isolation and eliminates the "double scheduler" overhead common in virtualized environments, while Tenon delivers precise, deterministic task management. By generating only the necessary software stack for each workload, TenonOS eliminates redundant layers, minimizes the Trusted Computing Base (TCB), and maximizes system responsiveness. Extensive evaluations demonstrate that TenonOS achieves superior real-time performance with a 40.28\% improvement in scheduling latency, maintains an ultra-compact memory footprint of 361 KiB, and exhibits high adaptability. These results validate TenonOS as an ideal foundation for next-generation, resource-constrained embedded systems requiring strict temporal guarantees.
\end{abstract}




\end{frontmatter}



\section{Introduction}
Operating systems (OSs) form the backbone of the information industry, providing the essential interface between hardware and software. At the foundational level, they manage and coordinate hardware resources; at a higher level, they support the development and evolution of complex application ecosystems~\cite{peterson1985operating, tanenbaum1997operating}. The enduring success of operating systems stems from their ability to offer robust, adaptable, and scalable environments for both developers and users. However, the rapid pace of technological innovation and the explosive growth of interconnected devices are increasingly straining traditional OS architectures~\cite{armbrust2010view, satyanarayanan2001pervasive}.

These limitations are particularly pronounced in the context of modern embedded environments, where applications are highly dynamic, heterogeneous, and resource constrained~\cite{satyanarayanan2017emergence, shi2016edge}. The proliferation of smart devices, sensors, and interconnected systems has resulted in a fragmented and diverse network landscape~\cite{gubbi2013internet}. Unlike traditional servers, these systems, ranging from industrial robots to autonomous vehicles, operate in close proximity to physical processes, demanding strict real-time responsiveness and deterministic behavior~\cite{shi2016edge}. Operating systems in these settings must manage diverse hardware platforms (e.g., multi-core SoCs with accelerators) while ensuring strong security and isolation under stringent power and memory constraints. Consequently, conventional OS architectures, originally designed for stable, centralized computing, often fail to accommodate the fragmented and time-sensitive nature of these modern workloads~\cite{zahran2019heterogeneous, zhang2024operating}.

A particularly pressing challenge in these complex environments is the management of mixed-criticality workloads. Many advanced embedded applications require distinct operating environments to coexist on a single device. For instance, autonomous vehicles commonly rely on a Real-Time Operating System (RTOS) to handle safety-critical control loops, while simultaneously running a General-Purpose Operating System (GPOS) for infotainment and high-level perception~\cite{wulf2021survey, mounir2019hardware, barletta2022achieving}. This architectural complexity underscores the growing difficulty in manually integrating and optimizing OS components. The demand for efficient coordination between these diverse subsystems highlights the urgent need for a framework capable of automatically tailoring the operating system structure to meet specific hardware and application requirements.

To address these needs, virtualization technologies have become a mainstream solution, enabling multiple isolated environments to run on shared hardware. By abstracting the underlying physical resources, virtualization improves flexibility and scalability, and provides strong isolation between different OS instances~\cite{shi2016edge, morabito2018consolidate}. However, virtualization is not without its drawbacks. Different layers—such as hypervisors, container runtimes, and guest operating systems—often employ independent resource management strategies, leading to inefficiencies. These inconsistencies can cause resource contention, increased latency, and degraded overall system performance, which are particularly problematic for latency-sensitive and resource-constrained applications~\cite{pan2019performance, sheikhalishahi2015autonomic, zhao2018cross, desina2023evaluating, krebs2015performance, mukherjee2015resource}. Moreover, the overhead and complexity introduced by hypervisor-based solutions may compromise real-time guarantees and make system design more challenging, especially when supporting a wide variety of hardware configurations.

To tackle these challenges, we propose \textbf{TenonOS}, a self-generating, intelligent, and lightweight operating system framework for time-critical embedded systems. TenonOS is built on a novel \textit{LibOS-on-LibOS} architecture that rethinks the conventional separation between the operating system and the hypervisor. Adopting a \textit{unified library-first} design, TenonOS decomposes both layers into modular micro-libraries, with each functionality implemented as a minimal, composable library. By supporting dynamic composition at runtime, TenonOS provides a flexible and efficient execution environment that can be tailored to different criticality levels and real-time requirements. This approach eliminates much of the redundancy in traditional OS–hypervisor stacks, enabling more compact, scalable, and hardware-aware deployments on resource-constrained platforms.

At the core of TenonOS is a two-layer architecture consisting of the hypervisor \textbf{Mortise} and the runtime \textbf{Tenon}. This structure is inspired by the natural growth of plants: Mortise serves as the fertile soil of the system, while Tenon acts as the seed from which multiple distinct operating systems can grow.

\textbf{Mortise} is a lightweight hypervisor that decomposes traditional virtualization functionality into modular micro-libraries. It supports efficient hardware resource management, low-latency inter-VM communication, and flexible dynamic allocation policies. \textbf{Tenon}, built atop Mortise, is a real-time capable LibOS designed to meet the timing constraints of modern embedded applications. Unlike conventional LibOS designs, Tenon integrates real-time scheduling mechanisms that support both cooperative and preemptive models, enabling deterministic task execution for critical workloads. It also provides multi-process support, preserving compatibility with general-purpose applications while maintaining a compact, modular structure. Mortise can concurrently host both Tenon and general-purpose operating systems, such as Linux, on the same hardware platform. This architecture supports mixed-criticality workloads and reduces the need for separate compute nodes.

A key feature of Mortise is its ability to manage the full lifecycle of Tenon instances, including on-demand instantiation, suspension, and termination of LibOS-based runtimes according to system workload and application requirements. By enabling fine-grained control over when and how operating system instances are created and destroyed, Mortise underpins TenonOS's self-generating capability. Rather than relying on static runtime environments, TenonOS can dynamically assemble, launch, and retire execution contexts as needed, allowing the system to elastically adapt to workload variations while minimizing resource consumption.

Together, \textbf{Mortise} and \textbf{Tenon} form a cohesive framework that bridges hypervisor-level resource control with application-level runtime flexibility. Mortise provides low-level isolation and scheduling, while Tenon delivers customizable execution environments optimized for responsiveness and scalability. This design aligns well with the diverse and performance-sensitive demands of resource-constrained and time-critical applications.

This paper presents the design and implementation of TenonOS, a generative operating system framework built upon a novel LibOS-on-LibOS architecture tailored for heterogeneous computing environments. Our main contributions are summarized as follows:

\begin{itemize}
\item \textbf{LibOS-on-LibOS system architecture.}
We propose TenonOS, a unified LibOS-on-LibOS architecture that
decomposes both the hypervisor and the operating system into a shared,
capability-based micro-library pool. This design collapses the traditional
OS--hypervisor layering and enables cross-layer code reuse and coherent
resource management, providing a foundation for time-critical embedded
deployments.

\item \textbf{Mortise: a minimal, dual-mode LibOS hypervisor.}
We design and implement Mortise, a LibOS-based type-1 hypervisor that
factorizes virtualization services into micro-libraries and supports both
static CPU partitioning and dynamic, hypervisor-driven scheduling modes.
Mortise provides strong isolation, low-overhead inter-VM communication,
and a substantially smaller trusted computing base than existing embedded
hypervisors.

\item \textbf{Tenon: a real-time, multi-process LibOS for time-critical workloads.}
We extend a cloud-oriented LibOS stack into Tenon, a real-time capable LibOS
that integrates a priority-based preemptive scheduler, fine-grained interrupt
handling, and lightweight multi-process support. Our experiments show that
Tenon achieves lower interrupt and scheduling latency than state-of-the-art
RTOSes such as Zephyr and RT-Thread, while preserving near bare-metal
behavior even when virtualized by Mortise.

\item \textbf{Open-source prototype and evaluation on ARM64 platforms.}
We provide an open-source implementation of TenonOS, including Tenon
(\url{https://gitee.com/tenonos/tenon.git}) and Mortise
(\url{https://gitee.com/tenonos/mortise.git}). Our prototype on ARM64
SoCs demonstrates a total code size of
11.3 K SLoC, a memory footprint of about 361 KiB, sub-40 ms Tenon
boot time, and negligible virtualization overhead on real-time metrics
even when co-located with Linux.
\end{itemize}

\section{Background and Motivation}

The landscape of embedded systems~\cite{hamdan2020edge,gamatie2009case} has evolved dramatically, transitioning from simple control tasks to complex, data-intensive workloads.
This distributed computing model brings computation and data storage closer to data sources, enabling real-time processing for applications such as autonomous vehicles~\cite{huang2023unmanned,zhou2020mobile,abrar2021energy}, industrial automation~\cite{stankovski2020impact,chalapathi2021industrial,weibin2019three}, and smart cities~\cite{khan2020edge,taleb2017mobile}. The unique characteristics of edge environments, including heterogeneous hardware architectures, resource-constrained devices, and geographically distributed deployments, pose significant challenges for traditional computing approaches.

The evolution of operating systems has historically been shaped by the needs of centralized computing environments. Classical operating system architectures, developed during the mainframe and personal-computing eras, were designed under assumptions of relatively homogeneous hardware and abundant resources~\cite{peterson1985operating,engler1995exokernel}. These systems typically adopt monolithic designs that abstract hardware resources through rigid interfaces. While effective in desktop settings, these architectures exhibit substantial limitations in embedded scenarios. Crucially, traditional monolithic kernels lack the flexibility to efficiently isolate diverse workloads—such as mixing real-time tasks with general-purpose computing—without introducing unpredictable interference and resource contention.


Modern embedded applications demand strict real-time guarantees and high-throughput responsiveness, necessitating effective workload consolidation~\cite{liu2019edge, mach2017mobile}. To meet these requirements, virtualization and containerization are commonly adopted to provide isolation and deployment flexibility~\cite{morabito2018consolidate}. However, these traditional solutions often incur prohibitive overheads in resource-constrained embedded environments. They typically rely on deep software stacks that involve trapping and emulating instructions or maintaining duplicate kernel structures~\cite{chiueh2005survey}. This architectural redundancy consumes valuable CPU cycles and memory, directly undermining the deterministic timing behavior required by critical tasks. Consequently, existing virtualization approaches struggle to strike the right balance between isolation and efficiency, highlighting the need for a lightweight framework capable of fine-grained, hardware-aware resource management.

\subsection{Issues with Monolithic Architectures}

\begin{figure*}
    \centering
    \includegraphics[width=1\linewidth]{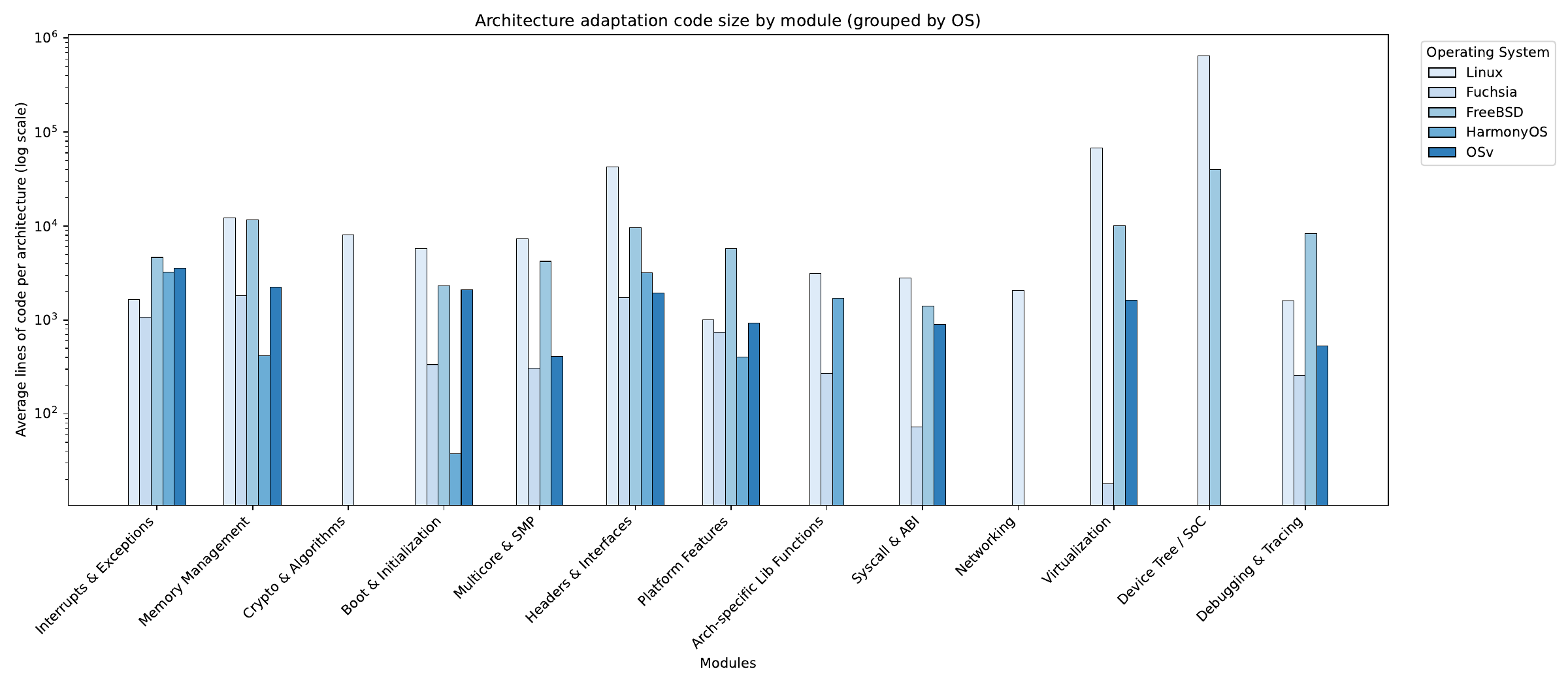}
    \caption{Average architecture-specific code size per module for each operating system . Bars are grouped by module on a logarithmic scale, showing how much architecture adaptation effort each OS concentrates in different subsystems.}
    \label{fig:adapt}
\end{figure*}

\begin{figure}[t]
    \centering

    \begin{subfigure}[t]{0.45\linewidth}
        \centering
        \includegraphics[width=\linewidth]{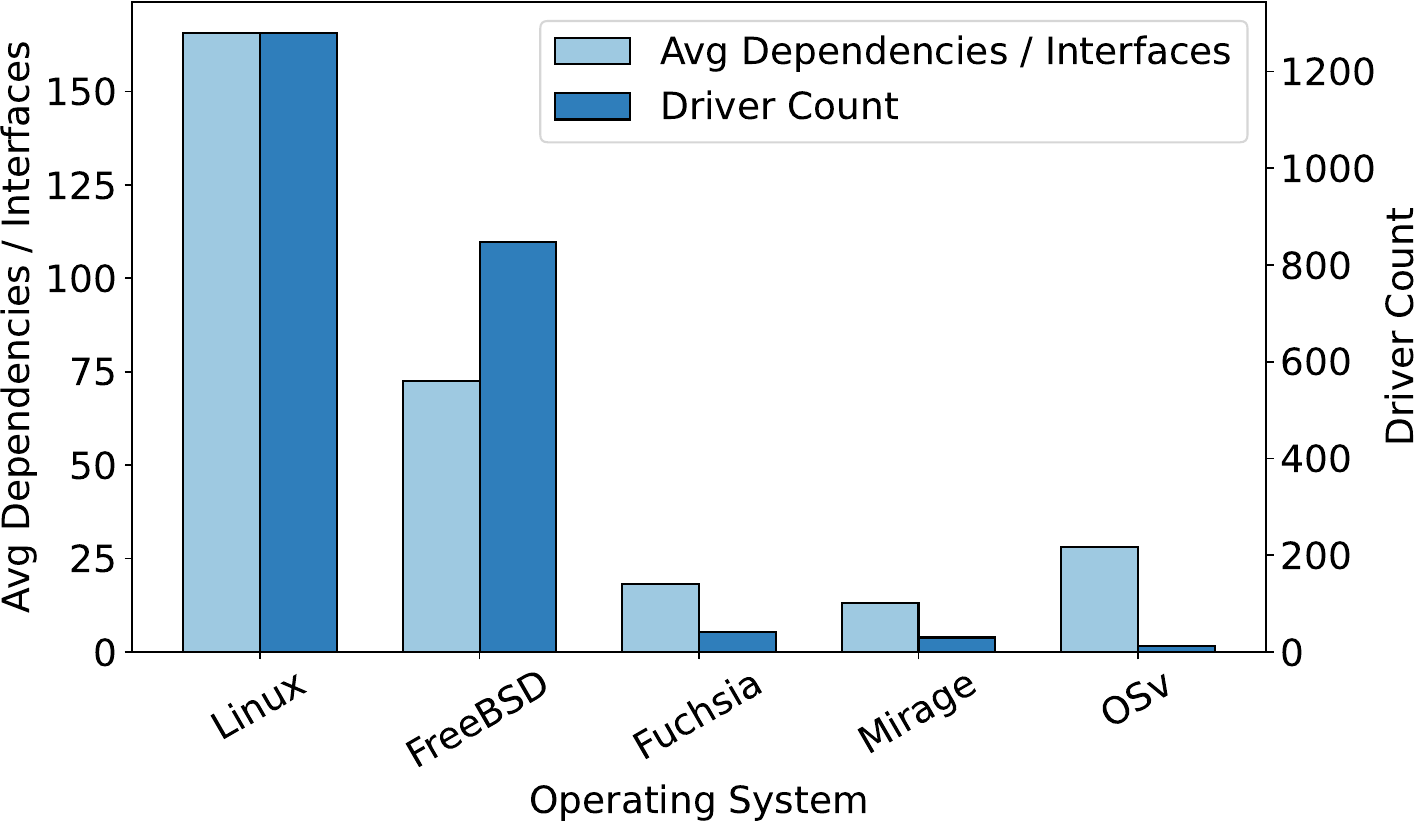}
        \caption{Average number of architecture-specific dependencies per driver (left axis) and total number of drivers (right axis) for each operating system.}
        \label{fig:deps_vs_drivers}
    \end{subfigure}
    \hfill
    \begin{subfigure}[t]{0.45\linewidth}
        \centering
        \includegraphics[width=\linewidth]{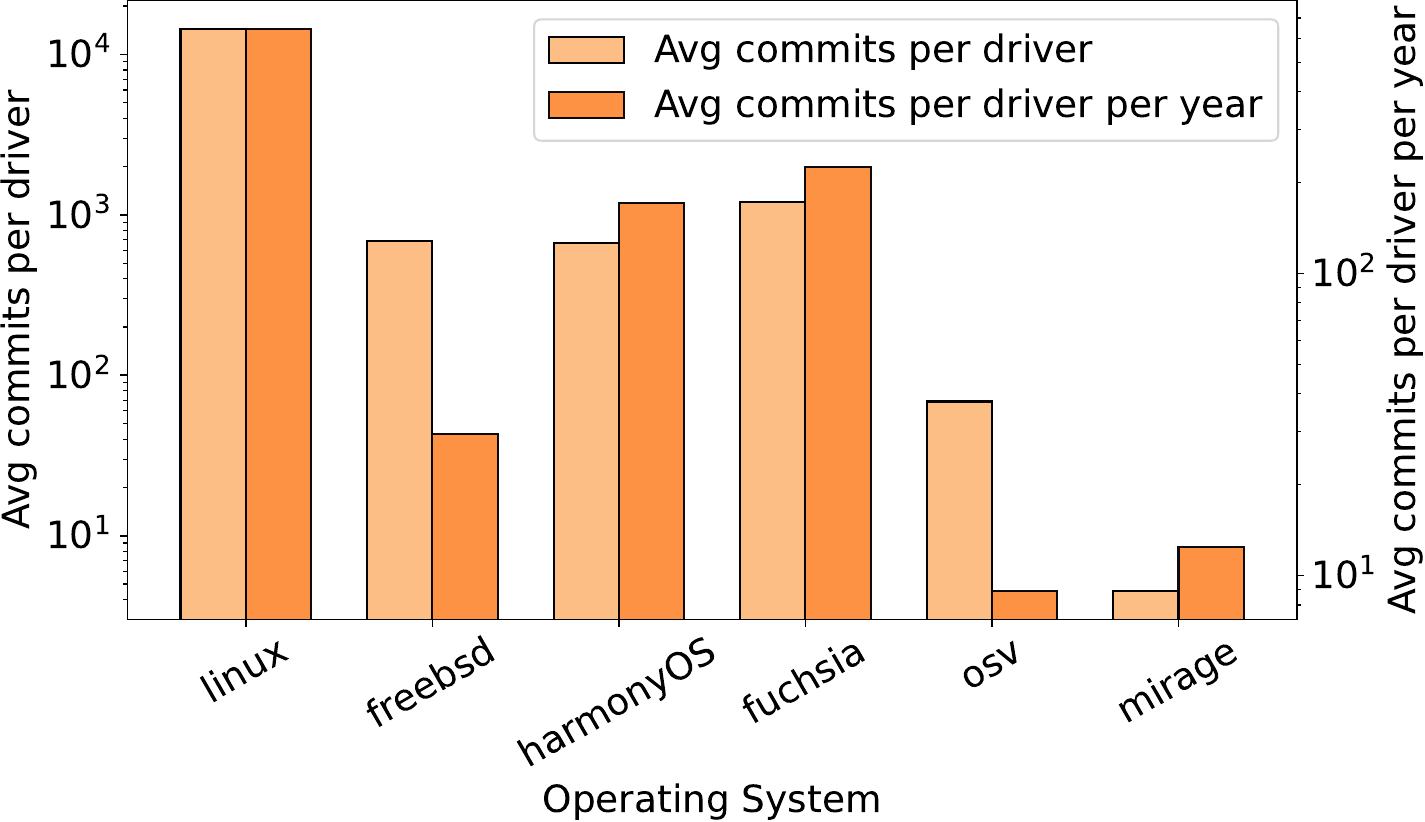}
        \caption{Average total commits per driver (left axis) and average yearly commits per driver (right axis, both in log scale) across operating systems.}
        \label{fig:commits_activity}
    \end{subfigure}

    \caption{Comparison of driver ecosystem size, architectural dependencies, and maintenance activity across different operating systems.}
    \label{fig:drivers_overview}
\end{figure}

In traditional monolithic operating system and virtualization architectures, all core functionalities are tightly coupled within a large and complex codebase. Although this design was adequate in earlier eras with relatively homogeneous and stable hardware, it exhibits significant limitations in today’s heterogeneous and rapidly evolving embedded environments. Figure~\ref{fig:adapt} shows that, across mainstream monolithic systems such as Linux and FreeBSD, architecture-specific code is spread over many subsystems and often reaches tens of thousands of lines per module. This broad and uneven distribution of adaptation code implies that supporting a new processor architecture or modifying an existing one requires invasive changes to multiple kernel components, increasing engineering effort and the probability of regressions. The driver ecosystem further amplifies these issues. As illustrated in Figure~\ref{fig:drivers_overview}, monolithic kernels maintain a very large number of drivers, each with many architecture-specific dependencies and frequent commits over time. This tight coupling between drivers and low-level architectural interfaces makes hardware evolution costly to support and complicates long-term maintenance, since even small architectural changes can trigger cascading updates across hundreds of drivers. Consequently, monolithic architectures struggle to provide efficient, low-risk evolution in the face of diverse accelerators, rapidly changing hardware platforms, and dynamic workloads, and are increasingly inadequate for modern embedded computing environments.

\subsection{The Case for Virtualization-OS Structures}
A key limitation stems from the layered architecture of traditional systems, where both the hypervisor and the OS independently manage hardware resources. This separation leads to duplicated functionality and conflicting policies. For example, virtual memory is maintained both at the hypervisor and guest OS levels~\cite{waldspurger2002memory}, and CPU scheduling occurs twice—first by the hypervisor for vCPUs, then by the OS for processes—resulting in possible preemption of critical tasks~\cite{menon2005diagnosing, govindan2007xen}. I/O requests also suffer from added latency due to cross-layer traversal~\cite{menon2006optimizing}, and debugging becomes more difficult due to poor visibility across layers~\cite{cherkasova2007virtual}.

Furthermore, many hypervisors rely on static or semi-static resource allocation~\cite{martins2020bao, guthrie2013vmware}, which cannot adapt to fluctuating workloads common in embedded scenarios. This often leads to underutilization or resource contention. Additionally, inter-VM communication—even between co-located guests—typically relies on virtual network stacks, introducing unnecessary overhead~\cite{barker2012empirical, ivanovic2018performance}.

These challenges reveal a deeper issue: the existing OS-hypervisor split lacks the architectural cohesion required to support flexible, low-overhead, and adaptive execution environments. Addressing this gap calls for a rethinking of how system software is structured—toward a model that merges resource management, reduces redundant abstraction, and supports diverse workloads with fine-grained control.

\subsection{Issues with Hypervisors in Embedded Environments}
Most current hypervisors adopt a monolithic architecture, which, while suitable for traditional server environments, introduces several limitations when deployed in embedded environments. Existing hypervisors are primarily designed for homogeneous hardware, focusing on uniform server-grade processors and devices. However, embedded environments are inherently heterogeneous, featuring a diverse mix of hardware accelerators such as GPUs, FPGAs, and specialized AI chips~\cite{ottaviano2024omnivisor,zha2021hetero}. Monolithic hypervisors struggle to efficiently manage and abstract these heterogeneous resources, making it difficult to extend support for new devices and often leading to suboptimal hardware utilization~\cite{gupta2015heterovisor,ottaviano2024omnivisor}.

Moreover, the increasing complexity of hypervisor codebases has made comprehensive security verification a significant challenge. Traditional monolithic type-1 hypervisors can comprise hundreds of thousands of lines of code~\cite{steinberg2010nova}, resulting in a large attack surface that is practically infeasible to fully audit or verify, especially for safety- and security-critical applications. This challenge is particularly pronounced in embedded environments, where devices frequently operate in physically untrusted or exposed environments, amplifying the risks associated with potential vulnerabilities in the hypervisor.

\section{Related Work}
Operating systems designed for embedded environments must carefully balance flexibility, performance, and strong isolation. Meeting these demands necessitates advancements in both operating system architecture and the underlying virtualization framework. Lightweight, application-specific operating systems—such as Library OSes (LibOS)—minimize overhead and enable precise control over resource usage. Concurrently, minimal type-1 hypervisors incorporating novel architectural designs provide robust isolation and predictable performance, both of which are critical for safety-critical and mixed-criticality applications.

\subsection{Type-1 Hypervisors}
Virtualization technology has advanced considerably, with type-1 hypervisors playing a key role in balancing performance, security, and resource isolation. Xen~\cite{barham2003xen}, a pioneer in type-1 hypervisor design, introduced the classic split-domain architecture. Running directly on bare-metal hardware, Xen separates device management (handled by Dom0) from guest virtual machines (hosted in DomU), ensuring strong isolation and security. Its support for both ARM and x86 architectures enhances compatibility in heterogeneous deployments. Extensions such as RT-Xen further enable latency-sensitive applications in industrial automation and IoT environments. Rust-Shyper~\cite{MO2023102948} represents a recent innovation at the programming language level. By leveraging Rust’s memory safety guarantees, Rust-Shyper minimizes vulnerabilities that are common in C/C++ implementations. This approach enhances security and reliability, which is especially important for embedded scenarios such as 5G base stations and distributed AI inference.

Further advancing hypervisor architecture, NOVA~\cite{steinberg2010nova} adopts a microhypervisor (microkernel-inspired) strategy, drastically reducing the trusted computing base (TCB) and improving system modularity. By delegating most functionalities to unprivileged user-space components, NOVA achieves strong fault tolerance, making it suitable for embedded systems deployed in untrusted environments like smart city infrastructure and autonomous devices. Continuing this trend towards modularity and determinism, BAO~\cite{martins2020bao} is a lightweight, bare-metal hypervisor designed for multi-core embedded systems. BAO applies a modular architecture and statically allocates hardware resources during initialization, achieving deterministic behavior with strong spatial and temporal isolation. This makes BAO particularly suitable for safety-critical and mixed-criticality applications, such as autonomous vehicles and medical devices.

This progression—from Xen’s classic architecture, through Rust-based language innovation, microkernel modularization in NOVA, to BAO’s lightweight and modular design—demonstrates how virtualization technology is evolving towards greater modularity, security, and efficiency for modern embedded computing demands.
\begin{figure*}[t]
    \centering
    \includegraphics[width=1\linewidth]{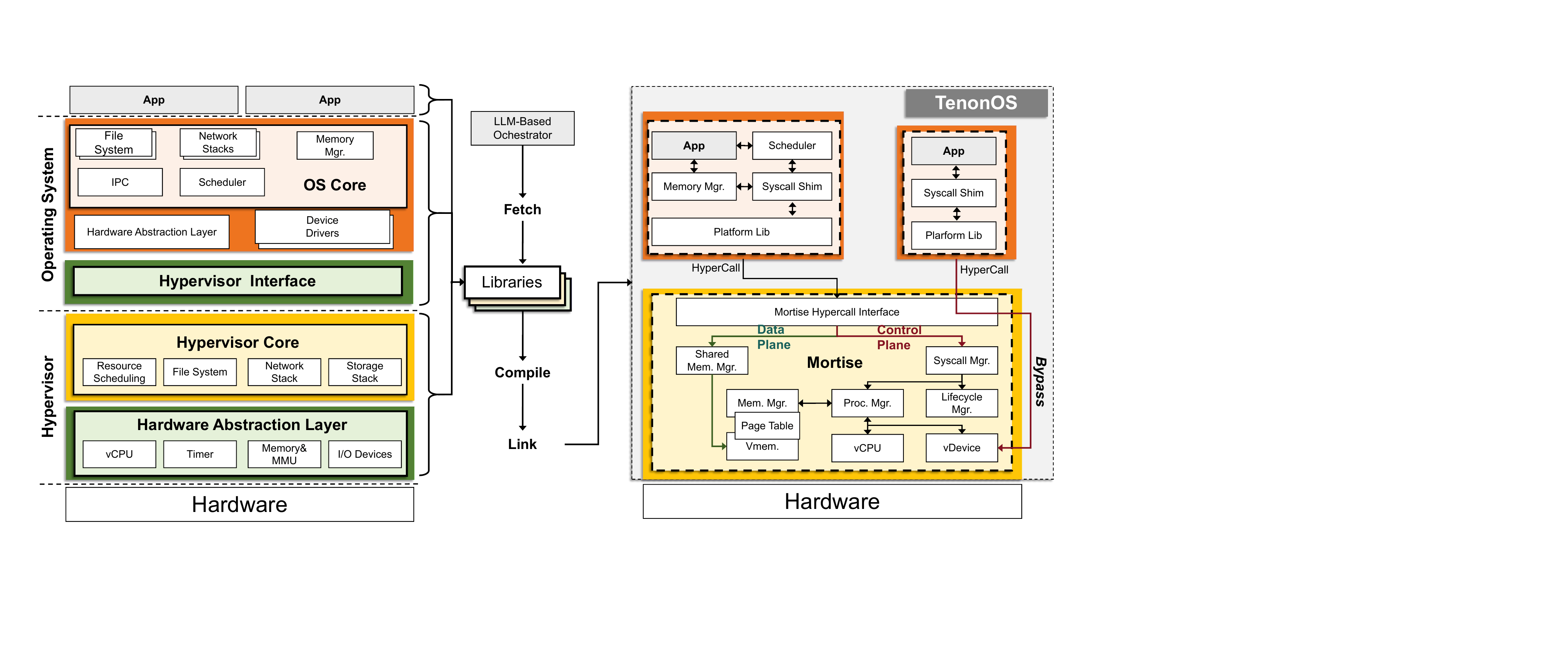}
    \caption{The overall architecture of TenonOS. The left section illustrates a traditional monolithic OS-Hypervisor design. The middle section depicts the \textit{Librarization} process, where system functionalities are modularized into a library pool comprising reusable components. The dynamic orchestration and composition mechanism allows rebuilding and generating TenonOS to adapt to heterogeneous hardware resources, forming a lightweight and agile LibOS on LibOS structure.}
    \label{fig:struc}
\end{figure*}

\subsection{Flexibility and Performance of Embedded LibOS}
LibOS  have emerged as a promising paradigm for constructing flexible, lightweight and high-performance software systems. Unlike traditional monolithic operating systems, a LibOS encapsulates only the essential OS abstractions and services required by an application, enabling application-specific optimizations. Early projects such as Exokernel~\cite{engler1998exokernel} pioneered the separation of application logic from resource management, allowing applications to directly manage hardware resources.

Recent advancements have focused on integrating LibOS into virtualized environments. For example, Graphene~\cite{tsai2017graphene} demonstrated the feasibility of running unmodified applications in isolated LibOS instances. This approach maintains compatibility with existing APIs while enhancing security and performance. In cloud computing, LibOS-based systems such as OSv~\cite{kivity2014osv} and Unikraft~\cite{kuenzer2021unikraft} have achieved significant improvements in startup latency, memory footprint, and runtime efficiency compared to traditional virtual machines. By leveraging their isolated architecture, these systems minimize overhead and enable further application-specific optimizations.

\section{System Design of TenonOS}
\subsection{TenonOS Architecture and Design Philosophy}
TenonOS is a generative operating system architecture characterized by the following key features:

\begin{itemize}
\item \textbf{Unified Library-First Design:} We reconstruct the traditional boundaries between the hypervisor and operating system by fully librarizing both layers. All functionalities are refactored into minimal, self-contained libraries, creating a unified, modular system stack that enables fine-grained customization and reuse.
\item \textbf{Dynamic LibOS Instantiation:} TenonOS supports on-demand generation of LibOS instances, enabling customized runtimes for specific applications, workloads and hardware.
\item \textbf{Code Reuse Across Stack:} The LibOS-on-LibOS model promotes code sharing between the OS and hypervisor, reducing maintenance effort and simplifying security auditing.
\item \textbf{Unified Resource Management:} By merging OS and hypervisor layers, TenonOS enables fine-grained, dynamic resource allocation across VMs, containers, and applications.
\item \textbf{Lightweight Inter-LibOS Communication:} TenonOS provides direct, efficient communication channels between LibOS instances, avoiding network-related overhead.
\end{itemize}

Figure~\ref{fig:struc} presents the overall architecture of \textbf{TenonOS}, which treats the virtualization layer and guest OS as a unified whole. Traditionally, as shown on the left of the figure, the hypervisor and operating system are organized in a monolithic, tightly coupled stack. In contrast, TenonOS proposes a novel \textit{Librarization} process, illustrated in the middle of the figure, where both the hypervisor and the OS are dismantled into a unified pool of reusable library components. These modules encapsulate core functionalities, and a dynamic library orchestration mechanism selects and composes the appropriate modules in response to specific hardware and application contexts. The right side of the figure shows the reconstructed modular stack: the bottom layer, \textbf{Mortise}, is a LibOS-based hypervisor that runs directly on hardware and can act as either a traditional OS or dynamically extend to a Type-1 hypervisor supporting multiple isolated runtimes. Above Mortise, the \textbf{Tenon} runtime adopts the same modular LibOS principles, enabling architectural consistency and seamless integration. Applications can either run within Tenon or, in some cases, bypass it to interact directly with Mortise for greater efficiency. This LibOS-on-LibOS structure enables flexible, fine-grained composition, rapid adaptation to diverse hardware, and improved maintainability and deployment portability.

\subsection{Shared Foundations and a Capability-Based Approach}
The virtualization layer and the operating system share foundational similarities in both functionality and purpose. Both manage critical hardware resources—such as CPU, memory, storage, and networking—and provide abstraction layers that shield developers and applications from low-level hardware details. They are designed to ensure isolation, preventing interference among applications, users, or virtual machines, and both incorporate security mechanisms to guard against unauthorized access and malware threats. Furthermore, they expose APIs and interfaces to facilitate application development and support multitasking, allowing multiple tasks (in operating systems) or virtual machines (in hypervisors) to run concurrently. These features are fundamental for efficient resource management and optimal hardware utilization.
Building on these shared foundations, we analyzed the codebases of traditional operating systems and hypervisors, and proposed a novel integration: a capability-based \textbf{micro-library} pool. In this approach, each function is modularized and provided as an independent micro-library associated with distinct capabilities. By encapsulating shared functionalities into a capability-based micro-library pool, the system achieves greater modularity, flexibility, and reusability. This design enables extensive code reuse across system components, reducing development effort, minimizing redundancy, and simplifying maintenance. Moreover, the strict association of micro-libraries with specific capabilities enhances security by limiting access to only the necessary functionality, thereby reducing the attack surface.

\subsection{Tenon}
To meet the growing demand for customizable and application-specific system designs, we introduce Tenon, a modular LibOS constructed using a flexible build tool. Tenon allows developers to add, remove, and recombine functional components, making it easier to adapt the system to various application scenarios. This modular approach enables the development of specialized operating systems suited for a wide range of workloads, from lightweight cloud services to complex embedded systems.
\begin{figure}
\centering
\includegraphics[width=1\linewidth]{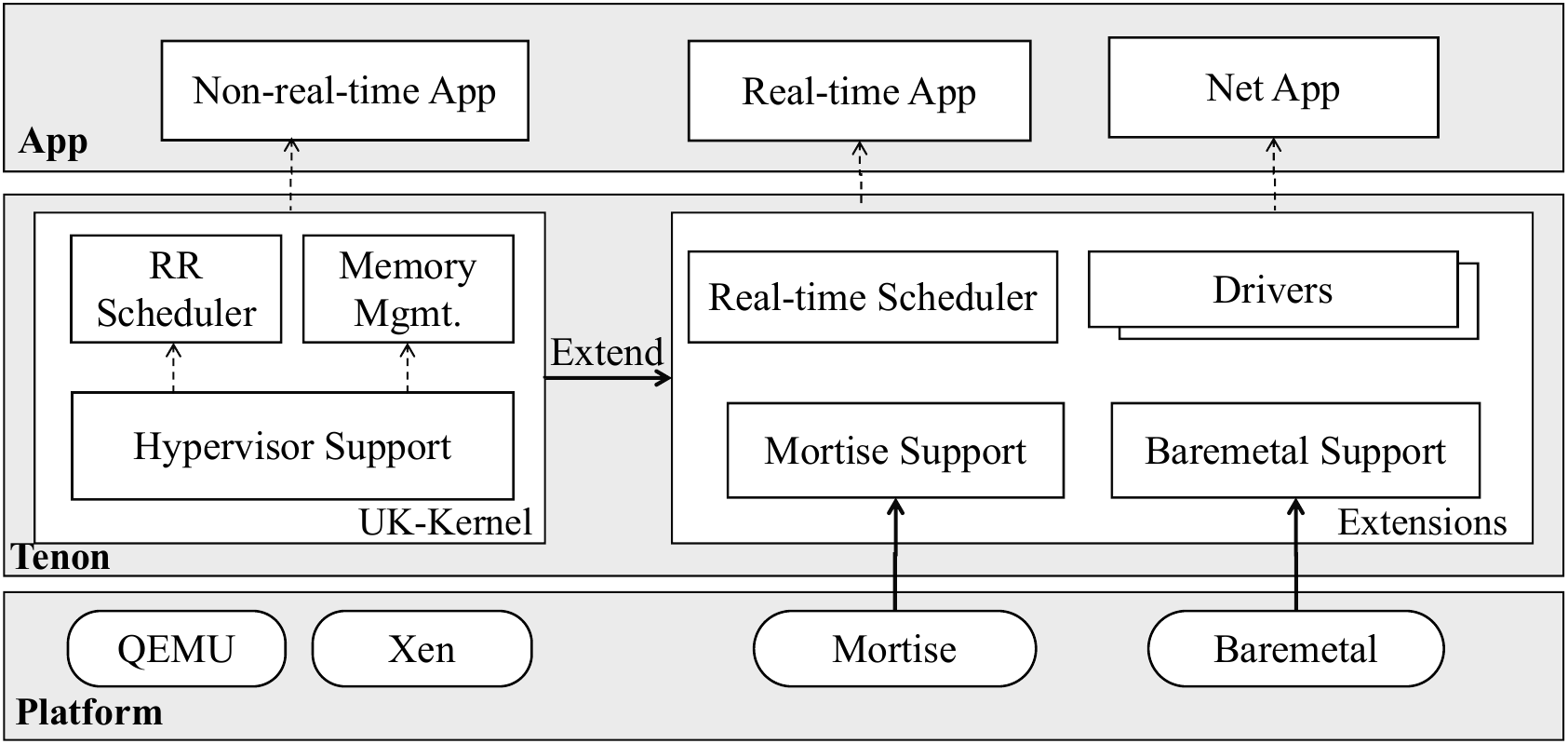}
\caption{Tenon extends UK-Libs (yellow) with kernel components (pink) such as a real-time scheduler, platform abstraction, and drivers, enabling real-time and networked applications across QEMU, Mortise, and bare-metal targets.}
\label{fig:tenonuk}
\end{figure}
LibOS implementations like Graphene~\cite{tsai2017graphene} and Unikraft~\cite{kuenzer2021unikraft} have demonstrated their effectiveness in providing lightweight environments for cloud computing. However, embedded systems—especially those in mixed-criticality environments or autonomous driving platforms—also require low-overhead and modular operating systems. Despite these needs, current LibOS designs lack real-time capabilities, which are essential for embedded applications. The absence of deterministic scheduling and reliable timing constraints makes them unsuitable for high-reliability use cases.

Another significant limitation is that most LibOS architectures rely on a single address space (SAS) and support only single-process execution. While this design simplifies implementation and enhances performance for single-application scenarios, it poses challenges when porting multi-process applications. Many general-purpose applications require process isolation, concurrency, and fault separation—features inherent to multi-process models. Adapting them to fit a single-process architecture often demands extensive refactoring, resulting in high migration costs and compatibility issues~\cite{peter2015arrakis}.

\begin{algorithm}
\setstretch{0.85} 

\caption{Thread scheduling flowchart illustrating preemption decisions, priority-based thread selection, context switching, and interrupt handling.}
\label{alg:scheduling}
\KwIn{Current thread \texttt{prev}}
\KwOut{Next running thread or end scheduling}

Disable interrupts\;
\eIf{Rescheduling needed and \texttt{prev} is ready}{
    \eIf{Preemption enabled and \texttt{prev} not locked}{
        \texttt{next} $\leftarrow$ highest-priority thread\;
        \eIf{\texttt{prev} $\neq$ \texttt{next}}{
            Enable interrupts\;
            Context switch (\texttt{prev}, \texttt{next})\;
            \Return
        }{
            Enable interrupts; \Return
        }
    }{
        Enable interrupts; \Return
    }
}{
    \eIf{\texttt{prev} not ready}{
        \texttt{next} $\leftarrow$ highest-priority thread\;
        \eIf{\texttt{prev} $\neq$ \texttt{next}}{
            Enable interrupts\;
            Context switch (\texttt{prev}, \texttt{next})\;
            \Return
        }{
            Enable interrupts; \Return
        }
    }{
        \eIf{\texttt{yield} and \texttt{prev} is ready}{
            \texttt{next} $\leftarrow$ highest-priority thread\;
            \eIf{\texttt{prev} $\neq$ \texttt{next}}{
                Enable interrupts\;
                Context switch (\texttt{prev}, \texttt{next})\;
                \Return
            }{
                Enable interrupts; \Return
            }
        }{
            Enable interrupts; \Return
        }
    }
}
\end{algorithm}

To address this, we introduce a lightweight process management system within Tenon, enabling support for multi-process execution. This enhancement improves compatibility with traditional applications and facilitates the deployment of more complex workloads.
Figure~\ref{fig:tenonuk} illustrates the architecture of Tenon across three representative configurations. The standard UK-Libs stack, which provides minimal OS abstractions such as memory management and I/O, supports single-process applications and can run in virtualized environments such as QEMU. Tenon extends this baseline by introducing two key kernel-level components: a real-time scheduler and a more flexible platform support layer. These additions enable fine-grained control over execution timing and hardware interaction, making the system suitable for real-time workloads. For more demanding network applications, Tenon incorporates a dedicated driver subsystem that interacts directly with hardware under a bare-metal setup. This modular architecture allows Tenon to operate consistently across different execution environments—QEMU, Mortise, and bare-metal—while preserving a unified programming interface.

As shown in Algorithm~\ref{alg:scheduling}, the scheduler uses a run queue organized by thread priority, ensuring that the highest-priority thread is always selected for execution. The resulting system is well-suited for embedded domains with strict timing requirements, such as autonomous vehicles and industrial control.

\subsection{Mortise}
Mortise draws inspiration from the LibOS design paradigm. Its architecture decomposes the conventional hypervisor into multiple independent micro-libraries, each responsible for a single function. These components can be developed, deployed, and verified separately, promoting modularity and maintainability.

\begin{figure}[t]
\centering
\includegraphics[width=\linewidth]{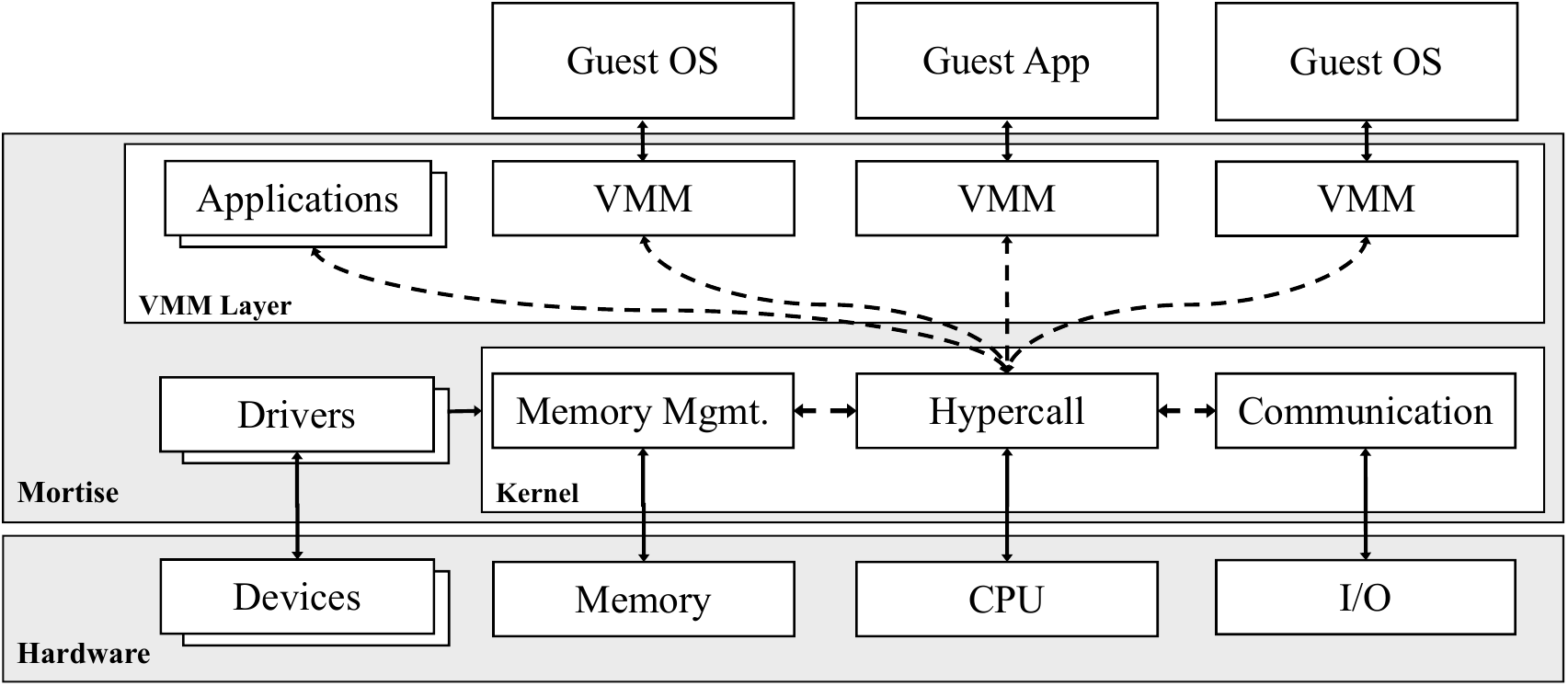}
\caption{Mortise architecture overview. Each vertical stack is a virtualized context managed by Mortise. The hypercall interface (yellow) provides guest and VMM access to privileged services like memory management and inter-VM communication. A shared driver layer (pink) mediates device access to physical hardware (blue)}
\label{fig:mortise-lib}
\end{figure}

Mortise adopts a LibOS-based architecture, as illustrated in Figure~\ref{fig:mortise-lib}. At its core, the Mortise kernel integrates essential OS functionalities, including memory management, scheduling, inter-process communication, and device drivers. These subsystems collectively manage hardware resources and provide unified, high-level abstractions to the upper layers. Mortise supports a variety of deployment models for both applications and guest operating systems: guest OSes can be executed via lightweight, isolated virtual machine monitors (VMMs), ensuring compatibility with legacy or specialized systems, while applications can be deployed natively on top of the kernel, benefitting from reduced overhead and enhanced security. This architecture enables the co-existence of multiple guest OSes and native applications within a unified runtime environment. By leveraging the LibOS model and VMMs, Mortise achieves strong resource isolation and robust security boundaries, while the single address space design eliminates the traditional kernel-user boundary, facilitating fast communication and efficient resource utilization. 

To meet the stringent performance and timing demands of time-critical workloads, Mortise supports executing applications directly on the virtualization layer, bypassing the guest OS entirely. Applications can be compiled with minimal runtime services and device drivers into a self-contained binary. This image runs directly atop the hypervisor without relying on a guest kernel, enabling fine-grained control over hardware and eliminating OS-level mediation.

\subsubsection{Hypercall Library}

The \textbf{hypercall} mechanism plays a vital role in virtualization by facilitating communication between guest virtual machines and the hypervisor. Similar to system calls in traditional operating systems, hypercalls allow VMs to request privileged operations that are otherwise restricted due to hardware-enforced isolation. These operations include memory allocation, CPU scheduling, interrupt injection, device I/O, and virtual machine lifecycle control such as pausing or restarting a VM.


Hypercalls in Mortise are registered and dispatched through a centralized routing mechanism. Developers group related hypercalls into function handlers using macros such as \texttt{REGISTER\_HYPERCALL\_GROUP}, which associate function IDs with corresponding logic. These handlers are organized into a routing table that maps incoming hypercall requests to their designated functions. When a guest issues a hypercall using architecture-specific instructions, control is transferred to the hypervisor, which consults the routing table to locate and invoke the appropriate handler. This modular design simplifies handler management, reduces dispatch overhead, and enables flexible extension of hypercall functionality.

\subsubsection{Guest OS Lifecycle Management Library}
Mortise enables lightweight lifecycle management of multiple operating system instances on a single host. Unlike traditional monolithic OSes that multiplex workloads through complex process scheduling, Mortise allows fine-grained OS instances to be created, paused, resumed, or terminated independently—each tailored for a specific function or workload.

To support efficient multi-OS coordination, Mortise introduces two scheduling modes. In \textbf{Mode 1 (Static Allocation)}, system resources such as CPUs and memory are pinned to specific OS instances, ensuring strong isolation and temporal predictability. In \textbf{Mode 2 (Dynamic Allocation)}, resources are shared and reassigned at runtime based on workload demands and system policies, allowing flexible adaptation to changing conditions.

This dual-mode scheduling design mitigates the inefficiencies of the “double scheduler” problem~\cite{rtxen}, where both the hypervisor and guest OS independently manage scheduling.

\begin{figure}[h]
\centering
\includegraphics[width=\linewidth]{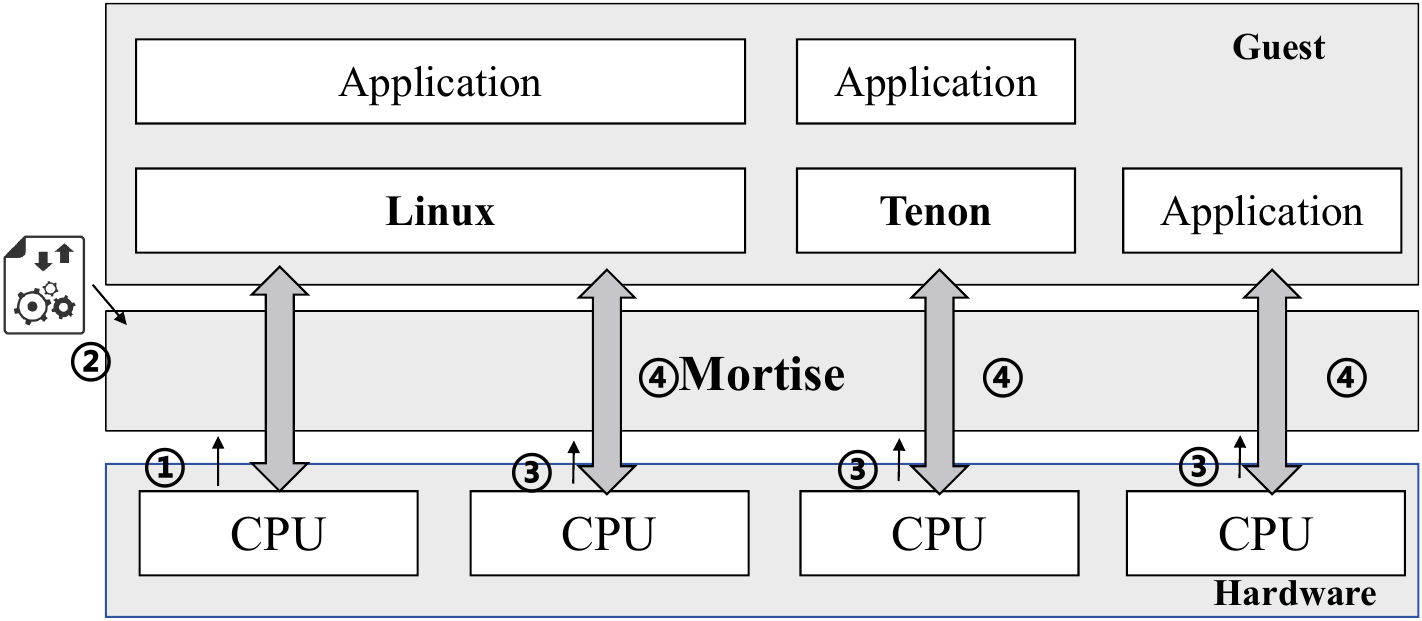}
\caption{\textbf{Mode 1 – Static Allocation Bootup:} 
\ding{172} The first CPU boots and initializes the Mortise hypervisor. 
\ding{173} Mortise reads a static configuration file that specifies the number of instances and their CPU assignments. 
\ding{174} All remaining CPUs register themselves with Mortise to participate in the system. 
\ding{175} Based on the configuration, Mortise launches Tenon or Linux instances on dedicated CPUs, ensuring strong hardware-level isolation between execution domains.}
\label{fig:mode1}
\end{figure}

\textbf{Mode 1} (Figure~\ref{fig:mode1}) enforces static CPU partitioning to ensure temporal and spatial isolation among LibOS instances. At boot, Mortise reads a configuration file defining the number of instances and their CPU assignments. Each instance is pinned to dedicated physical cores with no sharing, avoiding dynamic scheduling and enabling predictable execution.

Each CPU runs in its own address space, built with recursive page tables to support non-contiguous memory while minimizing page table walk overhead. Efficient TLB usage is maintained since CPUs do not switch address spaces.

System-level coordination uses selectively shared mappings, such as per-CPU buffers. Only CPUs within the same instance map its control structures, and hypervisor pages are protected with minimal, read-only or execute-only permissions. Hardware two-stage translation enforces further isolation. Superpages reduce TLB pressure, and guests handle interrupts and timers independently.

To minimize cache contention, Mortise optionally supports LLC partitioning via page coloring. Though this may increase memory fragmentation and startup latency, it can be enabled per instance based on need.

\begin{figure}[t]
\centering
\includegraphics[width=\linewidth]{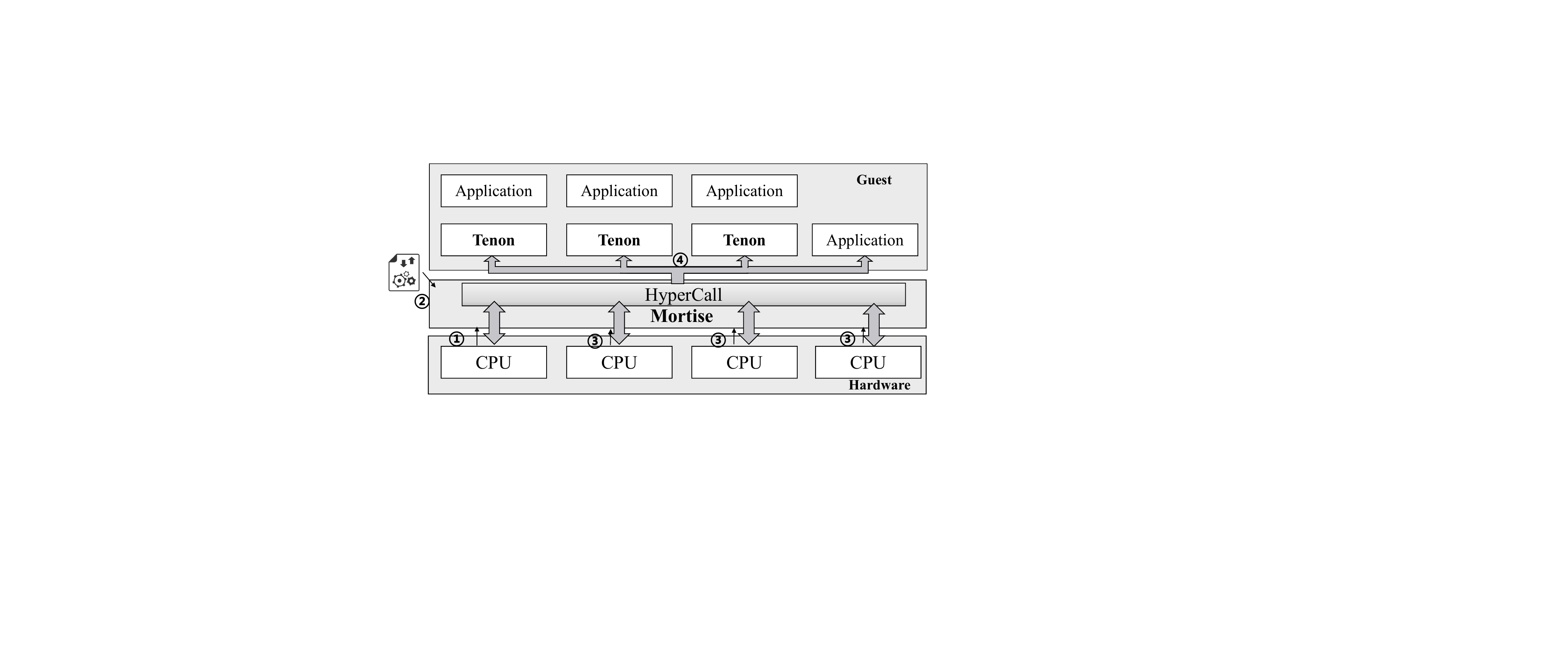}
\caption{\textbf{Mode 2 – Dynamic Allocation Bootup:} 
\ding{172} The first CPU boots and initializes the Mortise hypervisor. 
\ding{173} Mortise reads system configuration file that defines system policies. 
\ding{174} All CPUs register with Mortise, and hypercalls are enabled for dynamic instance control. 
\ding{175} Based on runtime requests via hypercalls, Mortise dynamically launches, pauses, or reclaims Tenon instances and assigns CPUs as needed.}
\label{fig:mode2}
\end{figure}

\textbf{Mode 2} (Figure~\ref{fig:mode2}) introduces a hypervisor-driven dynamic scheduling model. In this mode, the hypervisor centrally coordinates the execution of multiple LibOS instances, each managing a single process and relying on hypercalls for resource access and scheduling. The hypervisor schedules LibOS instances across CPU cores based on policies such as priority, workload type, and deadlines. Unlike the isolated static model, resources in dynamic mode are shared, enabling better utilization in environments with fluctuating workloads. Mortise supports preemptive scheduling, allowing high-priority tasks to interrupt lower-priority ones to meet timing constraints.

\subsubsection{OS Isolation and Communication}

Traditional hypervisors enforce strong isolation by assigning exclusive resources to each VM, which enhances security but complicates inter-VM communication. Due to strict separation, data exchange often relies on full network stacks, incurring high latency, overhead, and configuration complexity.

In contrast, intra-OS IPC benefits from shared kernel space, enabling low-latency mechanisms like shared memory or message queues with minimal setup. These approaches are efficient and developer-friendly, relying on standard APIs without virtualization-specific interfaces.

As multi-OS deployments become more common—particularly in embedded environments—efficient inter-VM communication (IVC) becomes essential. Scenarios involving real-time processing or dynamic service composition demand low-overhead channels that support both shared-memory and message-passing patterns. Mortise adopts a LibOS-inspired architecture that flattens the virtualization stack, enabling IVC mechanisms comparable to in-process IPC. By minimizing abstraction between guests and the hypervisor, Mortise supports lightweight, high-throughput communication between OS instances without sacrificing isolation, making it well-suited for collaborative workloads on constrained platforms.

\subsection{LLM-Based Dynamic Adaptation}
To support heterogeneous and dynamic embedded environments, TenonOS adopts an \textbf{LLM-based orchestration mechanism} that automatically selects and composes micro-libraries at runtime. This approach leverages the semantic reasoning capability of large language models to bridge the gap between high-level objectives and low-level system components, enabling demand-driven, application-specific runtime construction.  

The orchestration pipeline proceeds in three major stages:  

\begin{enumerate}
  \item \textbf{Objective Parsing}:  
  Natural language objectives provided by developers or system policies (e.g., ``optimize for real-time video analytics under power constraints'' or ``minimize memory consumption for lightweight IoT services'') are semantically parsed by the LLM. The output is a structured set of entities that capture performance goals, resource constraints, and security requirements.  

  \item \textbf{Graph-Guided Library Selection}:  
  To guide reasoning beyond plain text matching, TenonOS constructs a \textbf{Library Relation Graph (Lib-Graph)} that encodes the semantics and dependencies of micro-libraries. This graph is automatically built from Linux-style configuration metadata using a two-step process:  

  \begin{itemize}
    \item \textit{Entity Extraction}: Each \texttt{Kconfig} entry is parsed into an entity with attributes including the configuration name, type (e.g., \texttt{config} or \texttt{help\_text}), and textual description. Help text is included as a separate entity to preserve semantic context.  
    \item \textit{Relationship Extraction}: Hierarchical relations are derived directly from the Kconfig tree. Parent--child relations are added for nested options, and each config node is linked to its help text by a descriptive edge.  
  \end{itemize}

  The resulting representation is a directed acyclic graph where:
  \begin{itemize}
    \item \textbf{Nodes} represent configuration symbols and their associated help texts;  
    \item \textbf{Edges} encode semantic or hierarchical relations such as ``parent of'' or ``describes''.  
  \end{itemize}

  This graph is then loaded into the LightRAG framework, enabling semantic retrieval and consistency checks. During orchestration, the LLM explores paths in the Lib-Graph to identify relevant libraries. Each path is assigned a score based on relation semantics and contextual importance, and only those exceeding a predefined threshold are retained in the candidate library set.  

  \item \textbf{Configuration Generation and Validation}:  
  For each candidate library, the LLM infers appropriate parameter values (e.g., scheduling policies, memory allocation strategies, or I/O drivers). Proposed configurations are validated against the dependency rules embedded in the Lib-Graph, and invalid paths are pruned. A heuristic scoring mechanism then estimates the expected utility of each valid configuration for the stated objective, and the highest-scoring set is finalized.  
\end{enumerate}

Through this integration of \textbf{semantic orchestration} and \textbf{graph-based validation}, TenonOS can automatically assemble minimal yet sufficient runtime environments tailored to a given workload and hardware context. The graph ensures that only coherent and dependency-consistent libraries are composed, while the LLM provides flexibility in aligning abstract goals with concrete runtime components. This design significantly reduces redundant configurations, prevents invalid or hallucinated options, and enables rapid reconfiguration in embedded scenarios.

\section{Prototype}
This section presents the initial evaluation of TenonOS, focusing on performance and usability in multi-core embedded systems. We examine code size, memory usage, runtime overhead, and responsiveness—key metrics in resource-constrained settings requiring efficiency and predictability. The results highlight TenonOS’s practicality and scalability under real-world workloads.


\subsection{Overhead}

\subsubsection{Code Size and Memory Footprint}

\begin{table*}[t]
    \centering
    \caption{Summary of SLoC and memory usage for different components.}
    \begin{tabular}{lccc|ccccc}
        \toprule
        & \multicolumn{3}{c|}{\textbf{SLoC}} & \multicolumn{5}{c}{\textbf{size (bytes)}} \\
        \cmidrule(lr){2-4} \cmidrule(lr){5-9}
        & \textbf{C} & \textbf{asm} & \textbf{total} & \textbf{.text} & \textbf{.data} & \textbf{.bss} & \textbf{.rodata} & \textbf{total} \\
        \midrule
        arch/arm64              & 2354 & 454  & 2808  & 24648 & 3136  & 8852 & 0  &   36636\\
        platform  & 107  & 82    & 189   & 912   &  1484 & 16     & 0    &   2412  \\
        driver                    & 357 & 0    & 357  & 4644 & 992  & 206   & 0  & 5842  \\
        lib                     & 7994  & 0    & 7994   & 159732  & 24580    & 131959     & 48   &  316271  \\
        \midrule
        total                   & 10812 & 536   & 11348   & 189956  & 30192  & 141033  & 48  & 361229  \\
        \bottomrule
    \end{tabular}
    \label{tab:sloc_memory}
\end{table*}

We evaluate the code size and memory usage of TenonOS to assess its suitability for resource-constrained deployments. Table~\ref{tab:sloc_memory} summarizes the source lines of code (SLoC) and memory footprint across major components. The entire system comprises only 11,348 lines of code and occupies approximately 361 KiB in memory, including all text, data, and BSS sections.
\begin{table}[h]
    \centering
    \caption{Code size of shared micro-libraries reused across Mortise and Tenon}
    \renewcommand{\arraystretch}{1.5} 
    \begin{tabularx}{\columnwidth}{  
        >{\centering\arraybackslash}X
        >{\centering\arraybackslash}X
        >{\centering\arraybackslash}X
        >{\centering\arraybackslash}X
    }
        \toprule
        \textbf{Library} & \textbf{drivers/serial} & \textbf{lib/tntimer} & \textbf{lib/memory} \\
        \midrule
        Code size        & 357                     & 269                  & 519                 \\
        \bottomrule
    \end{tabularx}
    \label{tab:library_code_size}
\end{table}
The \texttt{arch/arm64} and \texttt{lib} directories constitute the bulk of the system, while platform-specific and driver components remain minimal. Table~\ref{tab:library_code_size} further illustrates the compactness of key libraries, such as serial I/O, timers, and memory management, each with fewer than 600 lines of code.

Compared to traditional hypervisors such as NOVA (20K LOC) and Bao (16K LOC), our virtualization layer is significantly smaller (11.3K LOC). This substantial reduction is largely attributed to the LibOS-based architecture, which delegates process and thread management to single-address-space LibOS instances rather than handling them within the hypervisor. By strictly limiting its scope to core virtualization functions—such as resource isolation and hypercall handling—TenonOS eliminates the need for complex subsystems typically found in monolithic hypervisors, including internal schedulers, IPC mechanisms, and device emulation layers. Consequently, this compact footprint simplifies verification and validation, minimizes the attack surface, and enables fast deployment, all of which are critical for safety-critical embedded environments.

\begin{table}[ht]
\centering
\caption{Boot performance metrics for different configurations.}
\resizebox{\columnwidth}{!}{%
\begin{tabular}{@{}l
                S[scientific-notation=true, round-mode=figures, round-precision=2]
                S[scientific-notation=true, round-mode=figures, round-precision=2]
                S[round-mode=places, round-precision=6]
                S[scientific-notation=true, round-mode=figures, round-precision=2]
                @{}}
\toprule
                       & \multicolumn{2}{c}{\textbf{hyp. init. time (s)}} & \multicolumn{2}{c}{\textbf{total boot time (s)}} \\
\cmidrule(lr){2-3} \cmidrule(lr){4-5}
\textbf{}              & {avg} & {std-dev} & {avg} & {std-dev} \\
\midrule
TenonOS bare                & {n/a}      & {n/a}      & 0.034564      & 4.0e-5    \\
TenonOS solo                & 2.54e-2    & 1.45e-5    & 0.0351792     & 1.28e-5   \\
TenonOS + Linux (co-locate) & 1.32e-1    & 3.04e-4    & 2.0467951     & 1.22e-3   \\
Linux bare                  & {n/a}      & {n/a}      & 1.8819968     & 5.61e-4   \\
Linux solo                  & 1.24e-1    & 3.20e-3    & 1.991333      & 2.86e-3   \\
Linux + Linux (co-locate)   & 1.86e-1    & 2.71e-3    & 2.4823773     & 9.42e-3   \\
\bottomrule
\end{tabular}%
}
\label{tab:boot_time}
\end{table}

\subsubsection{Boot Overhead}


The experimental results in Table \ref{tab:boot_time} highlight the lightweight nature of both Mortise and TenonOS. As shown in the boot time breakdown, TenonOS achieves extremely fast boot times on bare metal (0.0346 seconds) and when running as a solo guest in Mortise (0.00974 seconds), with negligible difference between the two. This demonstrates that the virtualization overhead introduced by Mortise is almost imperceptible for lightweight operating systems. Additionally, the Mortise hypervisor itself adds only a small and consistent initialization overhead (typically less than 0.2 seconds), further confirming its minimal footprint. Together, these results show that the combination of a lightweight hypervisor (Mortise) and a minimal OS (TenonOS) enables highly efficient and rapid system boot, making them well-suited for scenarios where fast startup and low resource consumption are critical.

\subsection{Real-time Efficiency}
Firstly, we evaluate Tenon with popular RTOS. 
\subsubsection{Baremetal Tenon Scheduling Performance}
\begin{table}[h]
    \centering
    \caption{Performance Comparison for Different RTOSes with Varying Thread Counts}
    \renewcommand{\arraystretch}{1.5} 
    \begin{tabular}{
        >{\centering\arraybackslash}p{5cm}
        >{\centering\arraybackslash}p{1cm}
        >{\centering\arraybackslash}p{1cm}
        >{\centering\arraybackslash}p{1cm}
        >{\centering\arraybackslash}p{1cm}
        >{\centering\arraybackslash}p{1cm}
    }
        \toprule
        \textbf{RTOS / Thread Count} & \textbf{2} & \textbf{10} & \textbf{20} & \textbf{50} & \textbf{100} \\
        \midrule
        Tenon (ectx enabled)   & 339 & 404 & 535 & 1392 & 2242 \\
        Tenon (ectx disabled)  & 166 & 242 & 304 & 708  & 1387 \\
        zephyr                 & 304 & 366 & 455 & 751  & 1380 \\
        \bottomrule
    \end{tabular}%
    \label{fig:same}
\end{table}

The experimental results in Table \ref{fig:same} demonstrate the effectiveness of Tenon in handling real-time scheduling scenarios. In the same-priority scheduling experiment, threads with identical priorities interact with the scheduler, achieving fair resource competition and maintaining stable scheduling latency. This highlights Tenon's ability to provide consistent and predictable scheduling in such scenarios. 


In contrast, the different-priority scheduling experiment in Table \ref{table:rtos_preemption} shows that high-priority threads can successfully preempt low-priority threads, as expected in a priority-based scheduling system. However, the response time of low-priority threads is noticeably affected, indicating potential challenges such as priority inversion or response time delays under heavy workloads. These results validate Tenon's capability to support priority-based preemption for real-time tasks while also revealing opportunities for further optimization to minimize response time fluctuations for lower-priority tasks.
\begin{table}[ht]
    \centering
    \caption{Preemption Times of Different RTOSes}
    \begin{tabular}{@{}lcccc@{}}
        \toprule
        \textbf{RTOS} & \textbf{preempt time (cycles)} \\
        \midrule
        Tenon with ectx    & 877  \\
        Tenon without ectx & 436  \\
        zephyr             & 656  \\
        rt-thread           & 1916 \\
        \bottomrule
    \end{tabular}
    \label{table:rtos_preemption}
\end{table}

\begin{figure}[h]
    \centering
    \includegraphics[width=0.95\linewidth]{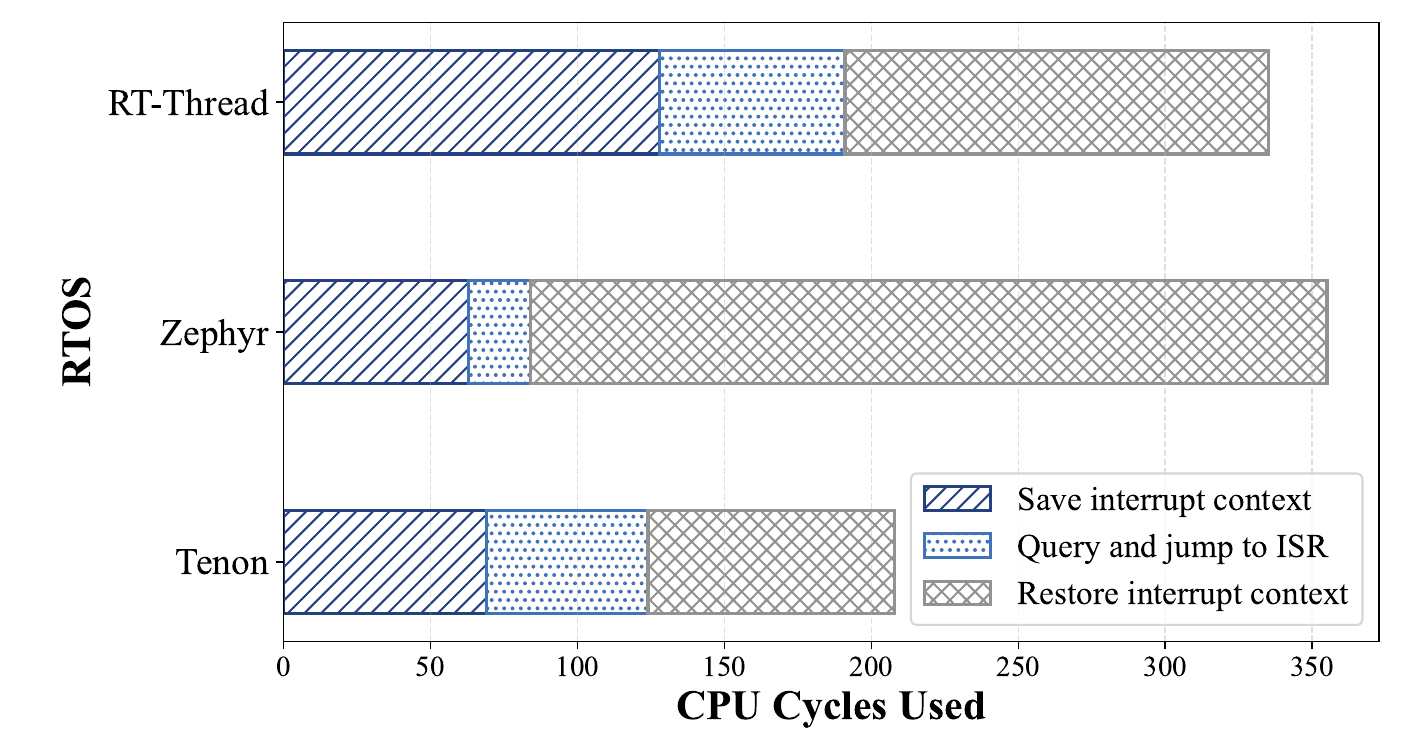}
    \caption{Performance Comparison of Different RTOS Phases}
    \label{fig:diff}
\end{figure}

Figure~\ref{fig:diff} shows a cycle-level comparison of interrupt handling across three RTOSs including Tenon, Zephyr, and RT-Thread, covering three phases: saving interrupt context, dispatching the Interrupt Service Routine (ISR), and restoring context. Tenon demonstrates the lowest total latency at 208 cycles, compared to RT-Thread (335 cycles) and Zephyr (355 cycles), while maintaining balanced performance across all phases. These results indicate that Tenon's interrupt model, which avoids reliance on generic kernel abstractions, is well-suited for time-sensitive workloads. Its predictable behavior makes it particularly appropriate for embedded systems requiring low and bounded latency.
\subsubsection{Baremetal Tenon Real-time Stability}

\begin{figure}[h]
    \centering
    \includegraphics[width=0.95\linewidth]{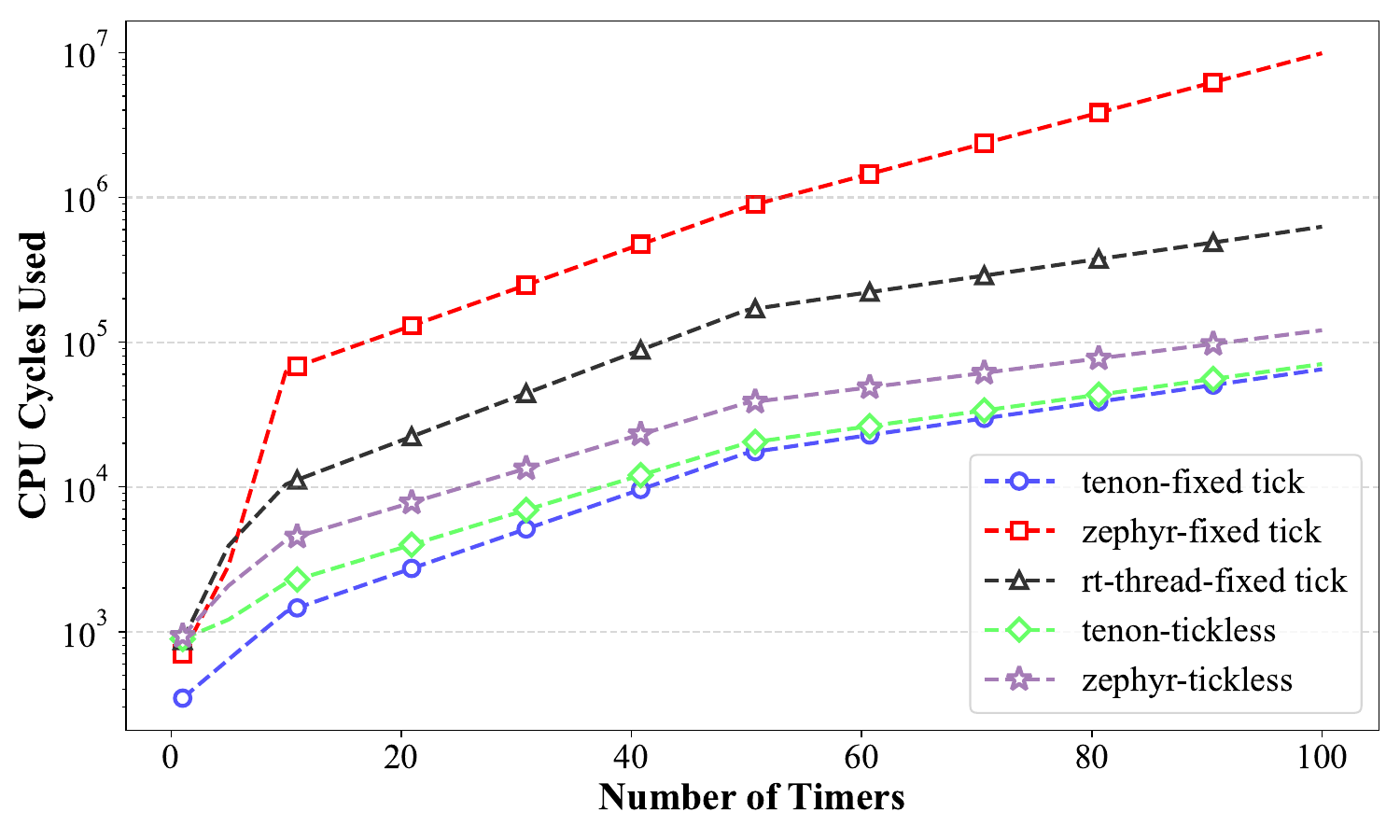}
    \caption{Performance Impact of Fixed Tick and Tickless Modes on Three RTOSes with Different Timer Counts}
    \label{fig:modes_combined}
\end{figure}

Figure \ref{fig:modes_combined} presents the computational cost as a function of the number of timers for various RTOSs and timer management strategies. The results clearly demonstrate the superior performance and scalability of Tenon, particularly in tickless mode.

As the number of timers increases, Tenon consistently achieves lower computational costs compared to other RTOS implementations. In fixed tick mode, Tenon shows moderate cost growth, remaining competitive with both Zephyr and RT-Thread. However, the advantage of Tenon becomes especially pronounced in tickless mode. The computational cost of Tenon-tickless grows very slowly as the number of timers increases, indicating excellent scalability and efficiency in handling large numbers of timers.
\begin{table*}[t]
    \centering
    \caption{Comparison of latency metrics (in CPU clock cycles) for the rk3568+TenonOS system under three configurations.}
    \renewcommand{\arraystretch}{1.2} 
    \small 

    \resizebox{\textwidth}{!}{%
    \begin{tabular}{@{}p{4.5cm}  
                    c c c c c@{}}
        \toprule
        \textbf{Environment / Test Item} 
            & \multicolumn{2}{c}{\textbf{Thread switches}} 
            & \textbf{Preemption} 
            & \textbf{Semaphore wakeup} 
            & \textbf{Interrupt} \\ 
        \cmidrule(lr){2-3}
        & \textbf{2} & \textbf{50} & & & \\ 
        \midrule
        Tenon                                 & 164 & 642 & 420 & 970 & 1290 \\ 
        Mortise + Tenon (2 cores)             & 164 & 648 & 420 & 970 & 1014 \\ 
        Mortise + Tenon (1 core) + Linux (2 cores) & 163 & 649 & 422 & 975 & 1051 \\ 
        \bottomrule
    \end{tabular}%
    }
    \label{tab:my_label}
\end{table*}
\subsection{TenonOS Running Efficiency}
\subsubsection{TenonOS Real-time Efficiency}
To evaluate the latency overhead introduced by \textbf{Mortise} and its effect on real-time performance, we conduct experiments under three configurations of the rk3568 platform running TenonOS. The results are presented in Table~\ref{tab:my_label}, with latency measured in \textbf{CPU cycles}.

\renewcommand{\arraystretch}{1.3} 
\setlength{\tabcolsep}{6pt} 

The results demonstrate the impact of integrating Mortise into the rk3568+TenonOS system across different configurations. For same-priority thread switches, there is negligible variation between the configurations, indicating that Mortise introduces minimal overhead in this scenario. Preemption and semaphore wakeup times remain largely consistent, with only minor differences observed between configurations, further supporting the efficiency of Mortise.

However, interrupt latency shows a noticeable improvement when Mortise is used. The Mortise + TenonOS (allocated 2 cores) configuration achieves the lowest interrupt latency (1014 cycles). The Mortise + TenonOS (1 core) + Linux (2 cores) configuration exhibits slightly higher interrupt latency (1051 cycles), likely due to resource sharing with Linux. Overall, these results highlight Mortise's ability to maintain real-time performance with minimal added overhead, even in scenarios involving concurrent Linux execution.

\subsection{LLM‑Based Orchestration}
\begin{figure}[t]
    \centering
    \includegraphics[width=1\linewidth]{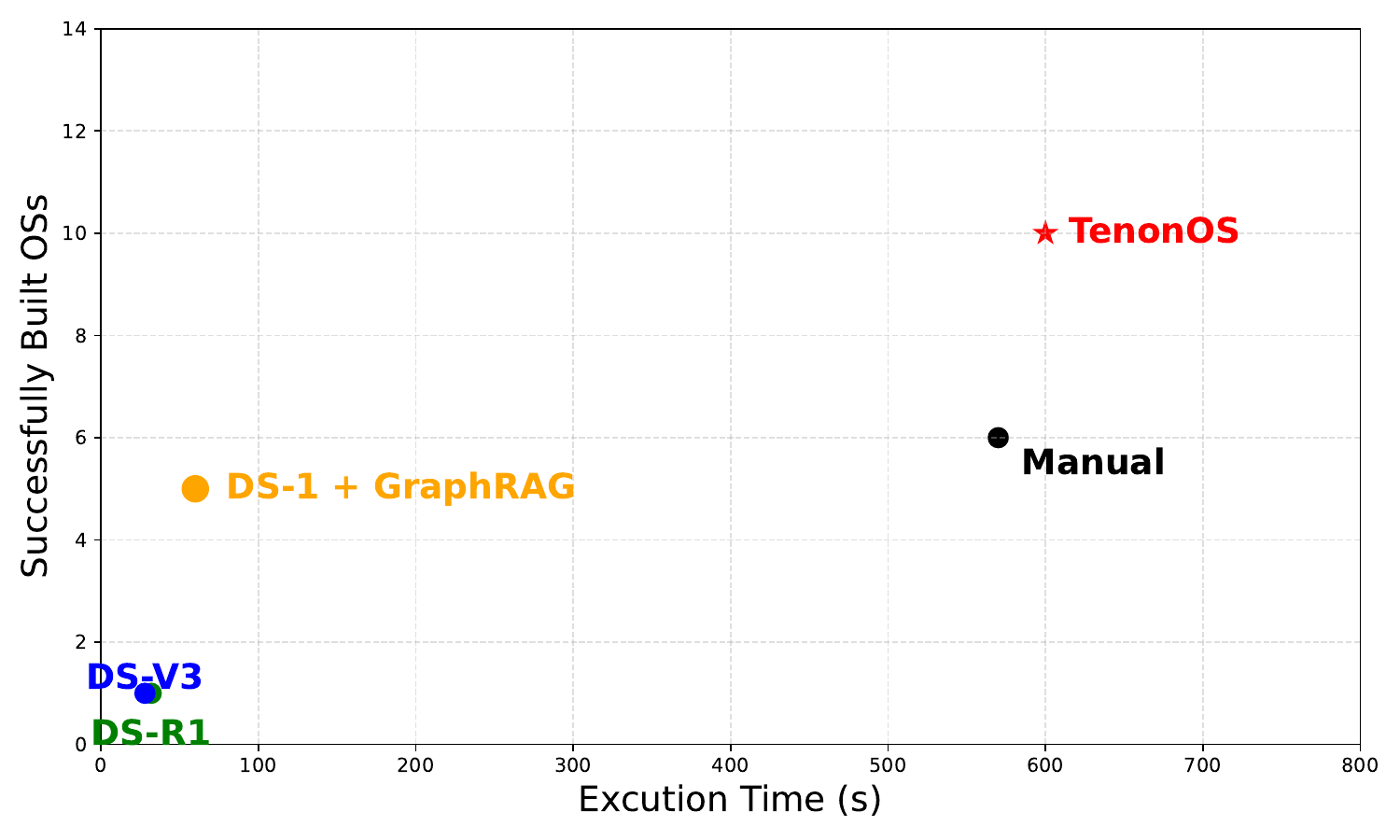}
    \caption{TenonOS generation efficiency and accuracy under different orchestration strategies. The x‑axis shows the end‑to‑end OS generation time, and the y‑axis shows the number of applications successfully built. TenonOS’s orchestration achieves the highest success rate with acceptable orchestration time, outperforming both pure LLM prompting and manual configuration.}
    \label{fig:orch}
\end{figure}

Figure~\ref{fig:orch} compares end‑to‑end generation time and build success for different orchestration strategies on a dataset of 14 Unikraft community–provided application configurations. Pure LLM prompting (DeepSeek‑R1, DeepSeek‑V3) is fast but rarely produces a working Unikraft configuration. Integrated into TenonOS, the orchestrator (TenonOS) achieves the best overall tradeoff, combining high build accuracy with acceptable orchestration latency.

\section{Conclusion}
In this work, we proposed TenonOS, a self-generating intelligent software framework for time-critical embedded systems that addresses the challenges of modern heterogeneous and dynamic computing environments. By combining lightweight LibOS principles with a modular architecture, TenonOS delivers a flexible and scalable solution capable of adapting to rapidly evolving software demands and diverse criticality levels. At its core, the Mortise hypervisor enhances system adaptability through efficient resource management, strong isolation, and dynamic CPU allocation, supporting both static and dynamic operational modes for co-located real-time and general-purpose workloads. Our evaluation demonstrates TenonOS's superior real-time performance, outperforming traditional RTOS solutions like Zephyr and RT-Thread in scheduling efficiency, latency predictability, and interrupt handling, while maintaining near bare-metal behavior even in mixed-criticality deployments where Tenon is virtualized by Mortise alongside Linux. The micro-library design of Mortise not only reduces the TCB for improved security but also enables high reusability and maintainability with minimal runtime overhead. Furthermore, TenonOS leverages AI-driven decision-making and a unified library pool to dynamically reconfigure itself across diverse hardware and software environments, from resource-constrained embedded SoCs to high-performance real-time applications. Collectively, these advancements position TenonOS as a next-generation RTOS framework that overcomes the limitations of monolithic architectures, offering a robust foundation for future heterogeneous and mixed-criticality computing systems.

\section{Future Work}

This work takes initial steps in building a modular micro-library system for TenonOS, but further efforts are needed to improve its scalability and hardware coverage. First, the current set of reusable micro-libraries remains small. Increasing library coverage will enable broader functionality and greater architectural flexibility. Future work will focus on identifying and refactoring additional components into independent, reusable modules. Second, TenonOS currently targets a narrow range of hardware platforms. To support heterogeneous deployment scenarios, we plan to expand compatibility across diverse processors, accelerators, and embedded devices. Addressing these areas will enhance TenonOS's adaptability and extend its applicability to a wider range of embedded and real-time systems.

\section*{Acknowledgments}
This work was supported in part by the fund of Laboratory for Advanced Computing and Intelligence Engineering, and in part by the National Science Foundation of China under Grants (62472375,62125206), and in part by the Major Program of National Natural Science Foundation of Zhejiang(LD24F020014, LD25F020002), and in part by the Zhejiang Pioneer (Jianbing) Project (2024C01032), and in part by the Ningbo Yongjiang Talent Programme(2023A-198-G).

\appendix

\bibliographystyle{elsarticle-num-names}   

\bibliography{main}

\end{document}